\documentclass[10pt,twocolumn]{article}
\usepackage{times}
\usepackage[dvips]{graphicx}
\usepackage[usenames,dvipsnames]{color}
\usepackage{amssymb}
\usepackage{amsmath}

\usepackage[T1]{fontenc}

\usepackage{color}
\usepackage{url}
\usepackage{hyperref}
\usepackage{apacite}

\begin{document}
\date{}
\title{\textbf{Usage Impact Factor: the effects of sample characteristics on usage-based impact metrics.}}
\author{\textbf{$\mbox{Johan Bollen}^{\star\dagger}$ and Herbert Van de Sompel$^{\dagger}$}\\
{\small {\it $^\dagger$ Digital Library Research \& Prototyping Team}}\\
{\small {Los Alamos National Laboratory}}\\
{\small \{ jbollen, herbertv\}@lanl.gov}\\
{\small LA-UR-06-7626}}
\maketitle{}

\begin{bf}
There exist ample demonstrations that indicators of scholarly impact analogous to the citation-based ISI Impact Factor can be derived from usage data. However, contrary to the ISI IF which is based on citation data generated by the global community of scholarly authors, so far usage can only be practically recorded at a local level leading to community-specific assessments of scholarly impact that are difficult to generalize to the global scholarly community. We define a journal Usage Impact Factor which mimics the definition of the Thomson Scientific's ISI Impact Factor. Usage Impact Factor rankings are calculated on the basis of a large-scale usage data set recorded for the California State University system from 2003 to 2005. The resulting journal rankings are then compared to Thomson Scientific's ISI Impact Factor which is used as a baseline indicator of general impact. Our results indicate that impact as derived from California State University usage reflects the particular scientific and demographic characteristics of its communities.
\end{bf}

\section{Introduction}

Usage of scholarly resources as recorded by digital information systems has been gaining acceptance as a tool to study the scholarly community. Usage data has been used to study trends in science \cite{detect:bollen2003} as well as to visually map the interests of certain subsets of the scholarly community \cite{mappin:bollen2006}. In addition, usage data has been shown to be a promising alternative to citation data in the assessment of scholarly impact. As early as 2001 \cite{readin:darmoni2002} propose a reading factor to rank journals according to their impact derived from a library's access statistics. \citeA{evalua:bollen2002} and \citeA{altern:bollen2005} propose the use of social network metrics calculated for journal networks derived from usage sequences in a library's access log. \citeA{worldw:kurtz2004,bibliom:kurtz2004} discuss the potential of usage data for impact ranking. \citeA{earlie:brody2006} later explore how early article usage statistics can predict citation rates.  In addition to these research developments, practical standards for publisher reported usage statistics (COUNTER project\footnote{\url{http://www.projectcounter.org/}}) and their aggregation (SUSHI project\footnote{\url{http://www.niso.org/committees/SUSHI/SUSHI_comm.html}}) have been developed. Thomson Scientific has recently included usage statistics in its ISI Web of Knowledge product\footnote{ISI Web of Knowledge Usage Reporting System (WURS)}.\\

Since usage data is recorded by particular information systems, the acquired data naturally pertains to the user community of those systems. For example, when \citeA{evalua:bollen2002} rank journals according to their usage this is done on the basis of usage data recorded by the Los Alamos National Laboratory Research Library servers and therefore reflects the preferences of the LANL community. In a similar manner, the results reported by \citeA{earlie:brody2006} apply to the user community of the UK arXiv mirror\footnote{\url{http://uk.arxiv.org/}}. A similar argument can be made for the "citation-download correlation tool" of the University of Southampton's CiteBase system\footnote{\url{http://www.citebase.org/}} which uses download information from the UK arXiv mirror.\\

In all cases the community for which usage was recorded is delimited by the boundaries of a particular service or information system. The resulting sample of the scholarly community that generated the usage data through its interaction with these systems is unknown both in terms of its diversity and span. The CiteBase user community could in fact be a diverse mix of undergraduate students, professors, university staff, laypersons, and scholars. Its span may or may not be limited to the United Kingdom. The resulting usage data and its subsequent analysis could therefore be shaped by a set of sample characteristics that are not well-understood. In fact, when considering usage statistics as a population statistic, the question then emerges for which sample of the scholarly community usage has been recorded, and how the characteristics of that particular sample will influence the outcomes of a subsequent assessment of scholarly impact based on these statistics.\\

The issue of sampling permeates the field of scholarly impact assessments, even where citation data is used. Thomson Scientific's ISI Impact Factor (ISI IF) is calculated from citation rates recorded for a set of ISI-selected journals. The corresponding sample of the scholarly community consequently has the following characteristics:
\begin{enumerate}
	\item {\em Span}: extends to the {\em global} set of scholarly authors.
	\item {\em Diversity}: limited to {\em scholarly authors}, and {\em articles} published in the set of ISI-selected journals.
\end{enumerate}

In spite of the latter limitation, the ISI IF is perceived to be based on a representative and respected sample which supports its general acceptance as an indicator of scholarly impact.\\

In comparison to the ISI IF, usage-based assessments of scholarly impact are generally based on samples of the scholarly community with the following characteristics:
\begin{enumerate}
	\item  {\em Span}: delimited by the {\em local} boundaries of a particular information service.
	\item  {\em Diversity}: extends to {\em all user types} who can request services for {\em any type of scholarly communication unit}.
\end{enumerate}

In order to realize impact measures derived from usage data that could achieve the same level of acceptance as the ISI IF, explorations along both the above dimensions need to take place. The first dimension, i.e. span, entails the aggregation of usage data across a wide range of services to create a more global, representative sample of the scholarly community, i.e. increase its {\em span}. In fact, \citeA{archit:bollen2006} propose an architecture for the large-scale aggregation of usage data which could be employed to achieve such global samples. This architecture however only addresses the technical issues involved in aggregating such samples; it does not address the issue of what constitutes a representative global sample, nor which services usage should be aggregated for. The second dimension, i.e. diversity, entails efforts to better understand and control how community characteristics, i.e. sample {\em diversity}, affect usage-based impact assessments, regardless of whether the sampled community is representative of the global scholarly community.\\

Whereas \cite{archit:bollen2006} is focused on aspects of the first dimension, i.e.~sample {\em span}, this article addresses the second dimension, i.e.~sample {\em diversity}: studying the effects of sample characteristics on usage-based assessments of impact. Usage of scholarly resources for all  23 California State University (CSU) campuses, comprising about 405,000 students and 44,000 faculty and staff, was recorded throughout the entire October 2003 to August 2005 period by the CSU linking servers \cite{openli:vandesompel2001}, thereby generating an extensive, high-granularity usage data set covering one of the world's largest and most diverse scholarly communities. A simple Usage Impact Factor (UIF) was defined to mimic the definition of the ISI IF and was then used to determine journal rankings on the basis of the recorded CSU usage data. Correlations between the resulting CSU UIF and ISI IF rankings are determined for a set of scholarly disciplines, demarcated by ISI journal classification codes. These correlations are then matched to the demographic features of the CSU community to yield insights into how they affect usage-based assessment of impact.

\section{Background}

\subsection{Citation Impact Factor}

The IF of a particular journal in a particular year as defined by \citeA{citati:garfield1979} is determined by counting the number of citations that occur in a given year to articles published in the journal during the two previous years and dividing that number by the total number of published items in that two year period. As such, the IF corresponds to the probability that the articles published in a particular journal over a 2 year period are effectively cited in a given year.\\

More formally, the IF can be defined as follows. We denote the set $A$ of (citable) articles published in journal $j$ in year $y$ as $A_j^y$ so that  $A_j^y = \{a_1, a_2, \cdots, a_n\}$, where $a_i \in A_j^y$ represents an article published in journal $j$ in year $y$. We introduce the citation function $C^y$ that maps a set of citable articles to the number of times these articles were cited by articles published in year $y$, i.e. $C^y(A) \rightarrow \mathbb{N}$. It follows that $C^y(A_j^{k})$ returns the number of citations recorded in year $y$ that point to the set of articles published in journal $j$ in year $k$.\\

The IF of a journal $j$ in year $y$, denoted $\mbox{IF}_j^y$, is defined as the ratio of two quantities:

\begin{equation}
	\mbox{IF}_j^y = \frac { C^y( A_j^{y-1} \cup A_j^{y-2} ) }   { | A_j^{y-1} \cup A_j^{y-2} | }
	\label{IFeq}
\end{equation}

where\\

$C^y( A_j^{y-1} \cup A_j^{y-2} )$ represents the number of citations in year $y$ to all citable articles published in journal $j$ in the two proceeding years $y-1$ and $y-2$,\\

and\\

$|A_j^{y-1} \cup A_j^{y-2}|$ represents the number of citable articles published by journal $j$ in the two proceeding years $y-1$ and $y-2$.

\subsection{Usage Impact Factor}

A similar reasoning can be applied to the definition  of a Usage Impact Factor (UIF) which can be framed in terms of the probability that an article published in a particular journal over a 2 year period is  {\em used}, rather than cited. Analogous to the IF, we define the Usage Impact Factor of journal $j$ in year $y$, denoted $\mbox{UIF}_j^y$, as follows. We replace the citation function $C^y(A_j^{k})$ with the usage function $R^y(A_j^{k}) \rightarrow \mathbb{N}$ which returns the number of times the articles in $A_j^{k}$ are used in year $y$. The UIF can then be defined as the ratio between two quantities:

\begin{equation}
\mbox{UIF}_j^y = \frac { R^y( A_j^{y-1} \cup A_j^{y-2} ) }   { | A_j^{y-1} \cup A_j^{y-2} | }
\label{UIFeq}
\end{equation}

where\\

$R^y( A_j^{y-1} \cup A_j^{y-2} )$ represents the number of uses recorded in year $y$ of articles published in journal $j$ in the two proceeding years $y-1$ and $y-2$ \\

and\\

$|A_j^{y-1} \cup A_j^{y-2}|$ represents the number of articles published by journal $j$ in the two proceeding years $y-1$ and $y-2$.\\

\begin{figure}[h!]
\begin{center}
\includegraphics[width=3in]{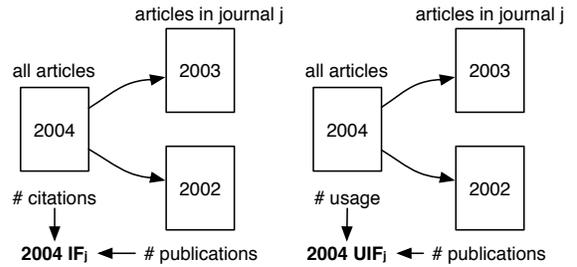}
\caption{\label{IF_v_UIF} Usage Impact Factor (UIF) defined in analogy to the ISI Impact Factor (ISI IF).}
\end{center}
\end{figure}

The UIF expresses the probability that an article published in a journal within a 2 year period is {\em used} in a particular year, much like the IF expresses the probability that an article published in a journal within a 2 year period is {\em cited} in a particular year. The similarities between the IF and the UIF are clarified in Fig. \ref{IF_v_UIF}.\\

To ensure that the IF and UIF for a particular journal are determined on the basis of similar samples, the UIF denominator $|A_j^{y-1} \cup A_j^{y-2}|$ is chosen to be that of the IF, namely the number of citable items published by journal $j$ in years $y-1$ and $y-2$. In other words, the number of citable or "usable" articles in a journal are considered the same quantity for a particular year.\\

In this work, we use the full-text downloads of an article as an approximation of article usage.  A similar problem of approximation exists in citation analysis where author motivations to cite a particular article can vary strongly \cite{proble:macroberts1989} and a citation can express any modality of agreement or interest. Contrary to citation data which lacks any formal indication of author motivation, usage logs typically do specify the user request type thereby allowing a careful selection of which to consider for a particular analysis. Although yet finer distinctions can be made between different types of usage, e.g. surveys to determine actual reading rates \cite{measur:king2006}, such an investigation was beyond the scope of this study; full-text downloads were considered to be the most reliable, if somewhat partial, indicator of usage.

\subsection{Data Acquisition}

\subsubsection{Sample Considerations}\label{sampleconsiderations}

The significance of sample span and diversity was outlined in the introduction. Therefore, when discussing usage- or citation-based metrics of impact, two orthogonal factors need to be taken into account :
\begin{enumerate}
	\item The characteristics of the sample that the specific metric has been calculated for, i.e.~sample span and diversity,
	\item The formal definition of a metric as an indicator of scholarly impact.
\end{enumerate}

\begin{figure}[h!]
\begin{center}
\includegraphics[width=1.4in]{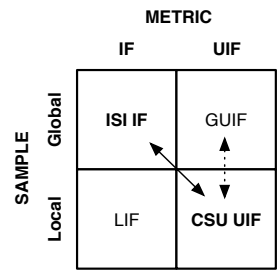}
\caption{\label{metric_ontology} Two orthogonal factors: formal metric definition and the sample to which it has been applied.}
\end{center}
\end{figure}

This perspective is represented in Fig. \ref{metric_ontology}. The IF, as defined in Eq. \ref{IFeq}, can be calculated for any set of journal citation data. However, the most common instantiation of the IF is the one published by Thomson Scientific's ISI. This ISI IF is calculated on the basis of citation data for a core set of about 8000 ISI-selected journals. With regards to the {\em span} of its sample, the ISI IF places no restrictions on the origin or affiliation of authors and therefore represents a {\em global} sample of the scholarly community, albeit one whose {\em diversity} is limited by the focus on authors who published journal articles in the set of ISI-selected journals.\\

The IF can be calculated for {\em local} citation samples. For example, \citeA{unders:mcdonald2006} extracts citation data pertaining only to California Institute of Technology authors to determine local citation impact. This approach results in a Local Impact Factor (LIF) as indicated in Fig. \ref{metric_ontology}.\\

The UIF as defined in Eq. \ref{UIFeq} can in principle be calculated for any usage data set, but the nature of usage data is such that it is generally recorded for the {\em local} user communities of a specific service.  This paper reports on UIF values calculated on the basis of usage data set for the California State University system which corresponds to a {\em local}, CSU-specific sample of the scholarly community. We therefore label the consequent UIF values "CSU UIF" to indicate the fact that they apply to {\em local} CSU usage.\\

The aggregation of usage data sets across different services and institutions may in the future yield increasingly global samples of the scholarly community. The resulting UIF rankings would then reflect a more global rather than a local, institutional sample of the scholarly community. Such metrics are labeled Global Usage Impact Factor (GUIF) in Fig. \ref{metric_ontology}.\\

This paper outlines a comparison of the globally oriented ISI IF which is used as a baseline indicator of general impact versus the CSU UIF which represents a local, CSU-specific facet of scholarly impact. It is however conceivable that once aggregated usage data becomes available a comparison between CSU UIF and the GUIF, the latter used as a global baseline, could be equally informative.

\subsubsection{ISI IF Citation Data}

ISI IF values were extracted from the 2004 Journal Citation Reports (JCR) that are published on a yearly basis by Thomson Scientific's ISI. Combined, the Science and Social Science edition of the 2004 JCR contained impact factors for 7,356 scholarly journals.

\subsubsection{CSU UIF Usage Data}

A large-scale usage log was created by aggregating usage data recorded by the linking servers \cite{refere1:vandesompel1999,refere2:vandesompel1999,openli:vandesompel2001} of the entire California State University system in the period October 2003 to August 2005.  Recording started November 11th, 2003 (10:44 AM) and continued uninterrupted until August 8th, 2005 (11:43PM). Linking server logs aggregate usage across all OpenURL-enabled information services, and thereby contain records of all user requests, including abstract requests and full-text downloads. They may additionally provide extensive usage, document and user metadata which allows e.g.~requester type, request types and publication dates to be taken into account when considering usage-based indicators of scholarly impact. As linking servers become increasingly prevalent, they achieve a growing importance among the tools by which enabled library services can record usage \cite{eviden:gallagher2005,unders:mcdonald2006}.\\

Usage for nine major institutions, i.e. Chancellor, California Polytechnic State University, CSU Los Angeles, CSU Northridge, CSU Sacramento, San Jose State University, CSU San Marcos, San Diego State University, and finally San Francisco State University was retained since they had recorded usage data most consistently and reliably, and represented the majority of CSU linking server data. A total of 3,679,325 unique usage events was thus recorded in the resulting master log for a total of 176,575 users (identified by their IP addresses\footnote{It is acknowledged that IP addresses do not uniquely identify individual users. However the presented analysis relies on overall article download frequencies and does not require  unique user identification.}), requesting services for 1,657,312 unique documents. A majority of the requests, i.e. 73\%, pertained to journal articles. A range of service request types was recorded, including but not limited to full-text downloads, requests for holding information, requests for journal citation data and abstract requests.\\

The resulting master log was then filtered to only include events conforming to the following:
\begin{enumerate}
	\item Article full-text downloads.
	\item Year of download was 2004.
	\item Download concerned articles published in 2002 and 2003.
\end{enumerate}

A total of 140,675 usage requests remained after this filtering. These events pertained to articles published in 6,423 unique journals. The number of full-text article downloads was tallied for each of these journals. The resulting download frequency table was then merged with the 2004 ISI IF data resulting in a list of 3,146 journals for which download data as well as non-zero ISI IF were available. Following Eq. \ref{UIFeq}, the journal download frequency values were then divided by the number of citable articles as was used to calculate the 2004 ISI IF, resulting in 2004 UIF values in conjunction with a 2004 ISI IF value for each journal.

\section{Results}

\subsection{CSU UIF journal rankings}

Table \ref{uif04-if04} lists the 10 journals with highest 2004 CSU UIF as well as their 2004 ISI IF values. The list of 10 journals with highest 2004 CSU UIF values reveals a  strong social science focus in the CSU community. The journals Topics in  Early Childhood Special Education (TOP EARLY CHILD SPEC), Hispanic Journal of Behavior Sciences (HISPANIC J BEHAV SCI), Intervention in School and Clinic (INTERV SCH CLIN) and Monographs of the Society for Research in Child Development (MONOGR SOC RES CHILD) are found at the top of the list. The low 2004 ISI IF values of these journals indicates a strong discrepancy between the degree by which journals are used by the CSU community and their overall scholarly impact as indicated by the 2004 ISI IF.\\

The 10 journals with highest 2004 ISI IF values are listed on the right-hand side of Table \ref{uif04-if04} along with their CSU UIF values. This ISI IF ranked list contains journals with high impact factor rankings such as Nature, Science, New England Journal of Medicine (NEW ENGL J MED), Cell and the Journal of the American Medical Association (JAMA). The corresponding 2004 CSU UIF values are relatively low for these journals in spite of their high 2004 ISI IF rankings.

\begin{table*}
\begin{center}
\begin{footnotesize}
\begin{tabular}{|l||l|l|l||l|l|l|}
		\multicolumn{4}{c}{{\bf Ordered by 2004 CSU UIF}}						&	\multicolumn{3}{c}{{\bf Ordered by 2004 ISI  IF}}	\\\hline
Rank	&	Title						&	UIF04	&	IF04		&	Title					&	UIF04		&	IF04		\\\hline
1		&	TOP EARLY CHILD SPEC	&	6.759	&	0.862	&	ANNU REV IMMUNOL	&	0.059		&	52.431	\\
2		&	HISPANIC J BEHAV SCI		&	6.720	&	0.500	&	CA-CANCER J CLIN	&	 0.667		&	44.515 	\\
3		&	INTERV SCH CLIN			&	6.017	&	0.172	&	NEW ENGL J MED		&	0.262		&	38.570	\\
4		&	MONOGR SOC RES CHILD	&	5.571	&	7.286	&	PHYSIOL REV			&	0.164		&	33.918	\\
5		&	J SCHOOL PSYCHOL		&	5.000	&	1.750	&	NATURE				&	0.277		&	32.182	\\
6		&	J FAM VIOLENCE			&	4.964	&	0.491	&	SCIENCE				&	0.288		&	31.853 	\\
7		&	SEX ROLES				&	4.804	&	0.639	&	ANNU REV BIOCHEM	&	0.077		&	31.538	\\
8		&	J YOUTH ADOLESCENCE	&	4.723	&	0.855	&	CELL				&	0.002		&	28.389	\\
9		&	EDUC URBAN SOC			&	4.653	&	0.224	&	JAMA-J AM MED ASSOC	&	1.196		&	24.831	\\
10		&	J AUTISM DEV DISORD		&	4.513	&	2.128	&	ANNU REV NEUROSCI 	&	0.048		&	23.143	\\\hline
\end{tabular}
\caption{\label{uif04-if04} Journals ranked by 2004 CSU UIF and 2004 ISI IF values.}
\end{footnotesize}
\end{center}
\end{table*}

\subsection{Correlating CSU UIF and the ISI IF}

\begin{figure}[h!]
\begin{center}
\includegraphics[width=3in]{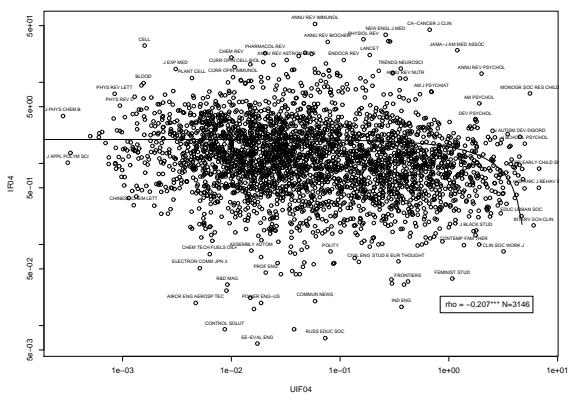}
\caption{\label{scatter_uif04-if04} CSU Usage Impact Factor and ISI Impact Factor values for 3,146 journals.}
\end{center}
\end{figure}

The Spearman rank order correlation coefficient between 2004 CSU UIF and 2004 ISI IF values was found to be -0.207 (p-value < 0.001, N=3,164) indicating a modest negative correlation between usage and the ISI IF for the  California State University community. This negative relationship is confirmed by the log-log scaled scatterplot in Fig. \ref{scatter_uif04-if04}. Some of the journals on the extremities of the scatterplot are labeled. It is notable that the journals with a high ISI IF value (top of plot), regardless of their 2004 CSU UIF values, mostly correspond to medicine. In addition, a significant number of prominent physics journals (Physical Review B and Physical Review Letters) are located in the quadrant of the plot which corresponds to high ISI IF and low CSU UIF values. In other words, they are considered high impact in the general scholarly community but their articles are used relatively infrequently in the CSU community.\\

This comparison of 2004 CSU UIF and 2004 ISI IF values fails to take into account variations among the different disciplines in the CSU system. A set of discipline-specific comparisons of the correlation between the 2004 CSU UIF and 2004 ISI IF is therefore provided in the following sections.

\subsection{Discipline-specific comparisons}

The scatterplot in Fig. \ref{scatter_uif04-if04} suggests that the relationship between the 2004 CSU UIF and 2004 ISI IF values differ for particular disciplines, e.g. among the set of journals with high ISI IF values and low CSU UIF values we find a preponderance of physics journals. It is therefore warranted to assess the CSU UIF and ISI IF  correlations within, rather than between, individual scholarly disciplines.\\

The disciplines used by CSU to tally enrollment and faculty numbers in its Statistical Abstracts \cite{csu:stats2004} are the starting point of the discipline-specific comparisons of 2004 CSU UIF and 2004 ISI IF values in this paper. These disciplines are listed in Table \ref{enroll} (reproduced from \citeA{csu:stats2004}, page 125, table 81).\\

\begin{table}[h!]
\begin{center}
\begin{footnotesize}
\begin{tabular}{||p{3in}||}
\hline\hline
{\bf Disciplines}		\\\hline\hline	                   			
Agriculture and Natural Resources, Architecture and Environmental Design, Area Studies, Biological Sciences, Business and Management, Communications, Computer and Information Sciences, Education, Engineering, Fine and Applied Arts, Foreign Languages, Health Professions, Home Economics, Interdisciplinary Studies, Letters, Library Science, Mathematics, Physical Sciences, Psychology, Public Affairs, Social Sciences\\\hline\hline					
\end{tabular}
\end{footnotesize}
\end{center}
\caption{\label{enroll} California State University disciplines used to tally enrollment and faculty numbers.}
\end{table}

\begin{figure}[h!]
\begin{center}
\includegraphics[width=3.1in]{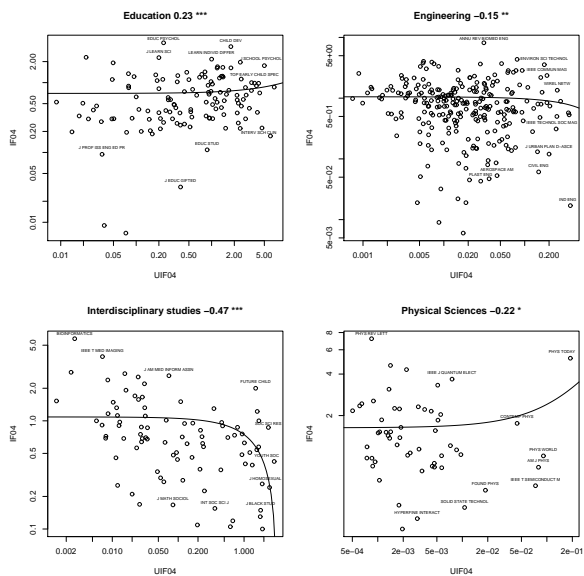}
\caption{\label{highcor_uif-if} CSU UIF and ISI IF comparisons for 4 disciplines with highest and lowest correlations.}
\end{center}
\end{figure}

To separate the group of examined journals in discipline-related sets, we manually matched each of the listed CSU disciplines with a set of  ISI journal classification codes\footnote{This is a subjective matter. However, specific care was taken to match ISI Journal Classification Codes as literally as possible to the specific CSU disciplines.}. These classification codes were then used to demarcate discipline-related sets of journals within which a comparison of CSU UIF and ISI IF could be conducted. The ISI journal classification codes for the CSU disciplines listed in Table \ref{enroll} are provided in Table \ref{categories} (appendix). The 2004 CSU UIF and 2004 ISI IF correlations calculated for each of the thus demarcated CSU disciplines are listed in Table \ref{domaincors}. Statistically significant correlations, marked in bold font, were found for only 3 of the 17 disciplines, namely Interdisciplinary Studies ($\rho=-0.470, N=89, p<0.001$), Education ($\rho=0.228, N=127, p=0.010$) and Engineering ($\rho=-0.147, N=259, p=0.018$). Physical Sciences  was found to have a marginally significant, negative correlation ($\rho=-0.225, N=56, p=0.096$). Log-log scaled scatterplots of the 2004 CSU UIF vs. 2004 ISI IF values for the mentioned four disciplines are shown in Fig. \ref{highcor_uif-if} and confirm the reported correlations.\\

\begin{table}[h!]
\begin{center}
\begin{footnotesize}
\begin{tabular}{||r|lll||}
							 \multicolumn{4}{r}{2004 CSU UIF vs. 2004 ISI IF}\\\hline
Discipline 					& 	rho				& 	N 	& 	p-value 	\\\hline
{\bf Interdisciplinary Studies} 		&	{\bf $-$0.470}			& 89		&	$>$0.001	\\
{\bf Education} 					&	{\bf $+$0.228}			& 127	&	0.010	\\
{\bf Engineering} 				&	{\bf $-$0.147}			& 259	&	0.018	\\
{\bf Physical Sciences} 			&	{\bf $-$0.225}			& 56		&	0.096	\\\hline
Agriculture and Natural Resources	&	$+$0.238				& 40		&	0.138	\\
Business and Management 		&	$+$0.132				& 115	&	0.160	 \\
Computer and Information Sciences &	$+$0.077				& 155	&	0.338	\\
Area Studies 					&	$+$0.169				& 27		&	0.397	 \\
Public Affairs 					&	$-$0.073				& 106	&	0.455	\\
Library 						&	$+$0.126				& 25		&	0.546	 \\
Psychology 					&	$+$0.033				& 316	&	0.556	\\
Architect. and Environ. Design 		&  $+$0.041				& 188	&	0.572	 \\
Mathematics 					&	$+$0.077				& 44		&	0.617	\\
Biological Sciences 				&	$-$0.024				& 331	&	0.669	\\
Communications				&	$+$0.049				& 58		&	0.712	 \\
Social Sciences				&	$+$0.026				& 59		&	0.843	\\
Health Professions				&	$-$0.012				& 126	&	0.890	 \\
\hline
\end{tabular}
\caption{\label{domaincors} Discipline-specific 2004 CSU UIF and 2004 ISI IF Spearman rank-order correlations.}
\end{footnotesize}
\end{center}
\end{table}

It is of particular interest that three out of the four mentioned disciplines exhibit a negative correlation between 2004 CSU UIF and 2004 ISI IF values. Whereas a zero correlation would have indicated the absence of a relationship, in this case the two metrics are inversely correlated indicating that members of the communities interested in the particular discipline specifically do not frequently use articles published in high-impact journals and vice versa. However, for Education a significant positive correlation was found between the 2004 CSU UIF and 2004 ISI IF, indicating that for this particular CSU discipline journal usage is moderately related to scholarly impact as indicated by the 2004 ISI IF.\\

The size of a discipline in terms of the number of journals that it comprises may affect ISI IF values. A marginally significant correlation was found between the CSU UIF and ISI IF correlation vs.  the number of journals in that particular discipline ($\rho=-0.459, N=17, p=0.065$). However, the correlation between CSU UIF and ISI IF values was not affected by the total number of students enrolled in a particular discipline. No statistically significant correlation was found between total student enrollment numbers and the correlation between CSU UIF and ISI IF correlations ($\rho=-0.262, N=17, p=0.308$).

\subsection{Community demographics}

On the basis of the hypothesis that the observed correlations between CSU UIF and ISI IF values for these disciplines may be related to the academic demographics of the CSU communities corresponding to the investigated disciplines, 2004 undergraduate and graduate enrollment and faculty numbers were matched to the observed correlations. Faculty numbers are estimated in terms of Full Time Equivalent Faculty (FTEF), i.e. the total number of hours taught in a particular division divided by the assumed 15 hours required for full-time faculty status. The particular number of FTEF and students respectively teaching or enrolled at the undergraduate or graduate level are listed in Table \ref{facstudents}. Note that undergraduate FTEF numbers are split into low and high divisions which need to be summed to determine total undergraduate FTEFs.\\

\begin{table}[h]
\begin{center}
\begin{scriptsize}
\begin{tabular}{||l||ll||ll|l||}
\hline\hline
								&	\multicolumn{2}{c||}{Students}	&	\multicolumn{3}{c||}{FTEF}				\\\hline
Discipline 	                   			& U.Grad 	& Grad.    		&	Low		&	High		&	Grad.	\\\hline		
Agri. \& Nat. Res.					& 5,381 	& 	302 		&	62.7		&	127.5	&	21.0		\\	
Arch. \& Env. Des. 	& 2,902 			& 	358 		&	33.9		&	72.1		&	19.2		\\
Area Studies						& 319 	& 	148 		&	12.9		&	25.1		&	4.3		\\
Biol. Sci.							& 13,642 	& 	1,052 	&	243.3	&	264.7	&	89.1		\\
Bus. \& Mngmt.						& 60,069 	& 	5,242 	&	143.3	&	914.4	&	161.3	\\
Communications					& 14,252 	& 	674		&	139.5	&	299.5	&	31.4		\\
Comp. \& Inf. Sci.     					& 16,415 	& 	2,322 	&	119.7	&	223.8	&	68.3		\\
{\bf Education}					& {\bf 16,084} 	& 	{\bf 15,452} 		&	{\bf 49.6}		&	{\bf 750.7}		&	{\bf 836.6}	\\
{\bf Engineering}				& {\bf 22,877} 	& 	{\bf 4,146} 		&	{\bf 191.8}		&	{\bf 483.6}		&	{\bf 123.9}	\\
Fine \& Appl. Arts					& 19,418 	& 	1,321 	&	425.3	&	712.1	&	102.0	\\
Foreign Lang.						& 2,252	& 	486		&	226.2	&	138.5	&	21.2		\\
Health Prof.						& 13,386	& 	3,984	&	31.2		&	142.9	&	143.1	\\
Home Econ.						& 3,261	& 	738		&	29.4		&	93.0		&	16.4		\\
{\bf Interdisc. Stud.}				& {\bf 29,780}	& 	{\bf 948}		&	{\bf 146.6} 		&	{\bf 225.5}		&	{\bf 24.8}		\\
Letters							& 13,594	& 	3,413	&	729.6	&	691.6	&	170.9	\\
Library							& -   		& 	561		&	6.6		&	2.0		&	17.3		\\	
Mathematics						& 3,325	& 	816		&	488.6	&	189.8	&	48.5		\\
{\bf Phys. Sci.}					& {\bf 3,310}	& 	{\bf 741}		&	{\bf 425.6}		&	{\bf 320.2}			&	{\bf 75.3}		\\
Psychology						& 16,944	& 	1,380	&	84.6		&	332.9	&	108.9	\\
Public Affairs						& 14,250	& 	4,643	&	47.4		&	287.0	&	216.8	\\
Social Sciences					& 24,597	& 	2,956	&	570.4	&	1,081.9	&	162.8	\\\hline\hline
\end{tabular}
\end{scriptsize}
\end{center}
\caption{\label{facstudents} California State University student enrollment and Full Time Equivalent Faculty (FTEF) numbers (undergraduate and graduate) for 2004.}
\end{table}

Three ratios of the undergraduate versus the graduate community were defined as follows:
\begin{enumerate}
	\item All: the ratio of total graduate student enrollment plus graduate FTEF numbers over the total number of undergraduate student enrollment plus undergraduate (high and low division combined) FTEF numbers.
	\item Student: the ratio of graduate over undergraduate student enrollment.
	\item Faculty: the ratio of graduate FTEF numbers over undergraduate FTEF numbers.
\end{enumerate}

The thus defined ratios were then compared to the observed CSU UIF vs. ISI IF correlations in Table \ref{domaincors}. It must be stressed this comparison was restricted to the mentioned four disciplines for which significant or marginally significant CSU UIF vs. ISI IF correlations were observed. The results are listed in Table \ref{ratio_cors} and suggest the possibility of a relationship between the ratio of the graduate to undergraduate community within a discipline and the observed CSU UIF vs. ISI IF correlations.\\

In particular, the discipline of Interdisciplinary Studies is characterized by a $\pm$15 to 1 ratio of undergraduate to graduate students, and a $\pm$ 30 to 1 ratio of undergraduate to graduate faculty. A highly significant negative  CSU UIF vs. ISI IF correlation was observed for this discipline.\\

Conversely, Education is characterized by a $\pm$1 to 1 ratio of undergraduate students and faculty to graduate students and faculty. A significant positive correlation was observed between journal CSU UIF vs. ISI IF values within this discipline.\\

This pattern is further confirmed by the undergraduate vs. graduate ratios for Engineering and Physical Sciences which has a moderate $\pm$5 to 1 and $\pm$10 to 1 undergraduate vs. graduate enrollment rate. Moderate negative  CSU UIF vs. ISI IF correlation were observed.\\

\begin{table*}
\begin{center}
\begin{footnotesize}
\begin{tabular}{l||rrr|rrr}
\multicolumn{4}{c}{}		&									\multicolumn{3}{r}{Grad. vs. Undergrad. ratio}				\\\hline
Discipline				&	$\rho$(UIF,IF) 	&	N	&	p-value	&	Student		&	Faculty 		&	All			\\\hline
Interdisciplinary Studies 	& 	{\bf $-$0.470}	&	89	&	0.000	&	{\bf 0.067}		&	{\bf 0.032}		&	{\bf 0.032} 	\\
Physical Sciences		&	$-$0.225		&	56	&	0.096	&	0.101		&	0.224		&	0.202		 \\
Engineering			&	$-$0.147		&	259	&	0.018	&	0.183		&	0.180		&	0.180		 \\
Education				&	{\bf $+$0.228}	&	127	&	0.010	&	{\bf 1.045}		&	{\bf 0.881} 	&	{\bf 0.888}		\\\hline
\end{tabular}
\caption{\label{ratio_cors} 2004 CSU UIF and ISI IF correlations compared to ratios of faculty and student numbers.}
\end{footnotesize}
\end{center}
\end{table*}

\begin{figure*}
\begin{center}
\includegraphics[width=6.5in]{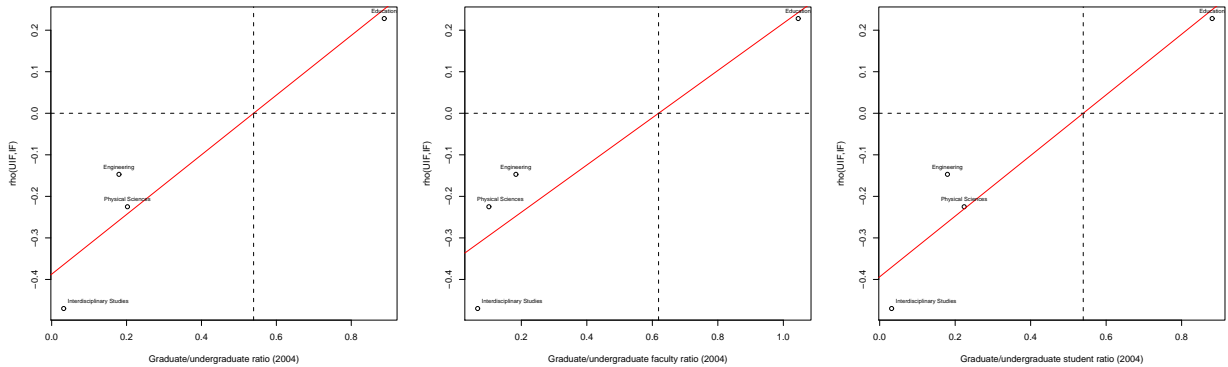}
\caption{\label{linmod_uif-if_enrol} Comparisons of Fall 2004 student and faculty populations vs.  2004 CSU UIF vs. ISI IF correlation.}
\end{center}
\end{figure*}

A linear regression model was generated for the relation between the ratio of graduate to undergraduate numbers versus the observed 2004 UIF and 2004 ISI IF correlation on the basis of the 4 data points listed in Table \ref{ratio_cors}. Since similar results were obtained for all three demographic ratios ("All", "Student and "Faculty"), only the linear regression model for the combined student and faculty ratios ("All") is discussed. Fig. \ref{linmod_uif-if_enrol} shows a scatterplot of the mentioned values and the corresponding linear regression model. The linear regression model was found to have an intercept of  -0.3873  and a slope of  0.7183 ($r^2=0.9029$).\\

From this it could be predicted that CSU UIF vs. ISI IF correlations become positive as soon as the graduate community becomes twice as large as the undergraduate community in a particular discipline. It can be noted that the overall ratio of graduate vs. undergraduate enrollment for the entire CSU system is 51,694 / 326,483 = 0.158, which together with the observed UIF vs. ISI IF correlation of $\rho = -0.207$ (p-value < 0.001, N=3,164) support the above mentioned pattern.

\subsection{Baseline assessment}

The 2004 ISI IF is used as a baseline assessment of scholarly impact against which 2004 CSU UIF values can be compared. Although CSU UIF and ISI IF are deliberately compared for the same years in which usage, citation and publication samples were recorded, questions arise with regards to the sensitivity of this comparison to longitudinal changes in the ISI IF over time.\\

\begin{figure}[h!]
\begin{center}
\includegraphics[width=3in]{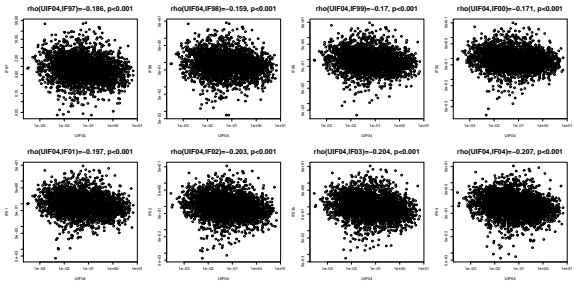}
\caption{\label{uif-if_8years} CSU UIF vs. ISI IF comparisons for 1997-2004 period.}
\end{center}
\end{figure}

For this reason we investigated the degree of correlation between the 2004 CSU UIF vs. past ISI IF values, i.e. ISI IF values that were published in 1997 through 2004\footnote{At the time this analysis was conducted, 2005 ISI IF values were not yet available.}. The results are listed Table \ref{uif-if97-04tbl}. These correlations indicate a stable, negative correlation between 2004 CSU UIF values and past ISI IF values over the mentioned period of 8 years. The absence of a particular trend in  CSU UIF vs. ISI IF correlations is supported by the plot in Table \ref{uif-if97-04tbl}. The scatterplots of CSU UIF vs. ISI IF values for each specific year are shown in Fig. \ref{uif-if_8years}.

\begin{table*}
\begin{footnotesize}
\begin{center}
\begin{tabular}{r|cccccccc}
\multicolumn{9}{c}{ \includegraphics[width=4in]{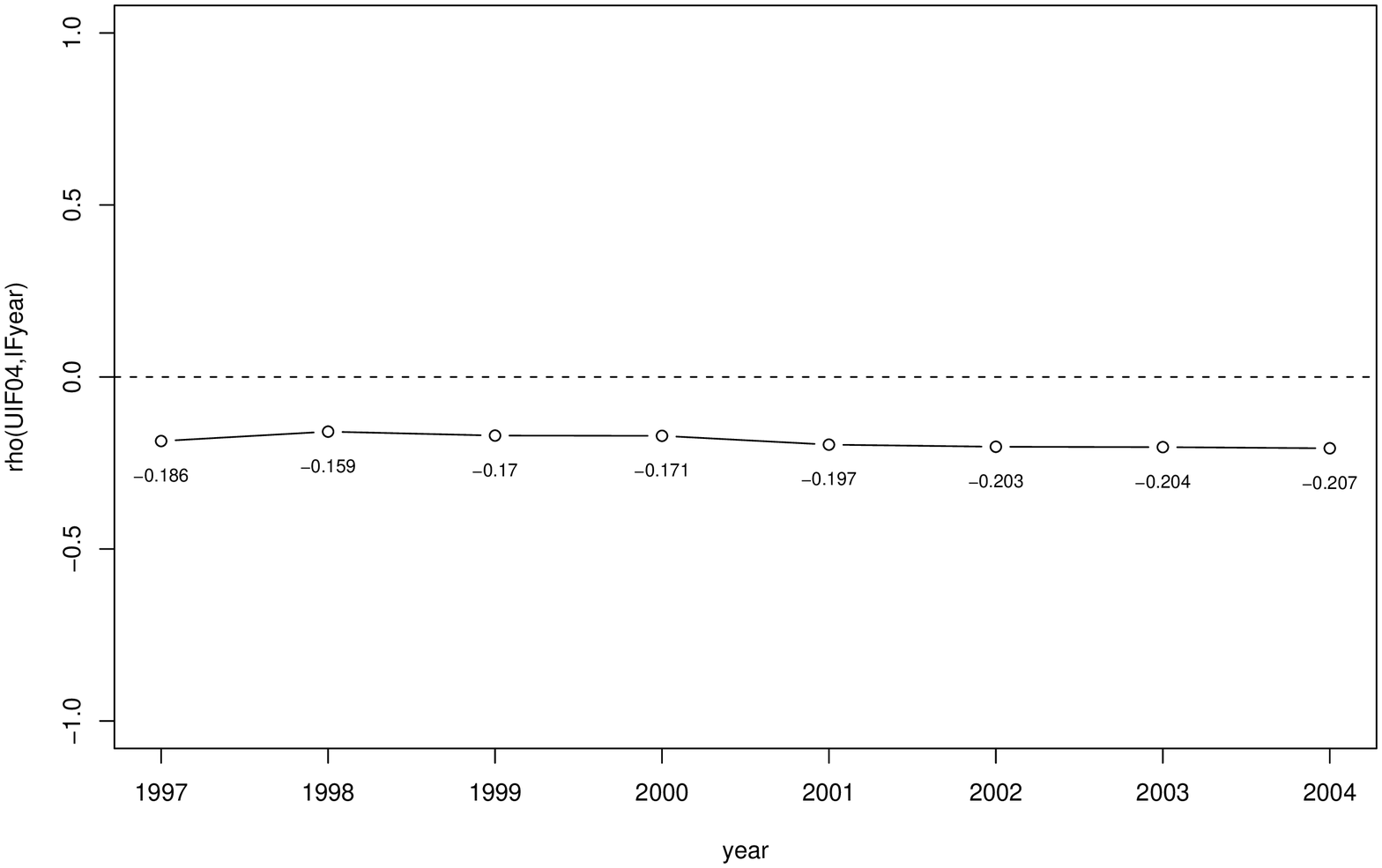}}\\
\hline\hline
ISI IF year			&	ISI IF 1997		&	ISI IF 1998		&	ISI IF 1999		&	ISI IF 2000		&	ISI IF 2001		&	ISI IF 2002		&	ISI IF 2003		&	ISI IF 2004\\\hline
2004 CSU UIF		&	 {\bf $-$0.186}		&	{\bf $-$0.159}		&	{\bf $-$0.170}	&	{\bf $-$0.171}	&	{\bf $-$0.197}	&	{\bf $-$0.203}	&	{\bf $-$0.204}	&	{\bf $-$0.207}\\
N				&	2636			&	2750			&	2819		&	2892		&	2960		&	3050		&	3096		&	3146 \\
p-value			&	<0.001		&	<0.001		&	<0.001	&	<0.001	&	<0.001	&	<0.001	&	<0.001	&	<0.001\\\hline
\hline
\end{tabular}
\end{center}
\end{footnotesize}
\caption{\label{uif-if97-04tbl} Spearman rank-order correlation values between 2004 Usage Impact Factor and 1997-2004 ISI ISI Impact Factors.}
\end{table*}

\subsection{Results Summary}

The picture that emerges from these results can be summarized as follows:

\begin{enumerate}
	\item A moderate negative correlation between 2004 CSU UIF and 2004  ISI IF values was found without taking into account CSU disciplines.
	\item This negative correlation persists over a period of 8 years counting back ISI IF values from the year in which usage was recorded (2004). 
	\item Some CSU disciplines exhibit negative correlations between CSU UIF and ISI IF values whereas others exhibit
		positive correlations. Most disciplines however exhibit zero or insignificant correlations.
	\item CSU UIF vs. ISI IF correlations seemed to be related to the ratio between the sizes of the undergraduate and graduate community in a discipline.
\end{enumerate}

\section{Conclusion}

Usage-based metrics of scholarly impact are gradually gaining acceptance in the domain of bibliometrics. However, little attention has been paid to how usage-based impact assessments are influenced by the demographic and scholarly characteristics of particular communities. The discussed analysis of CSU usage data indicates significant, community-based deviations between local usage impact and global citation impact as indicated by the generated CSU UIF and ISI IF rankings respectively. In particular, we found a general negative correlation between the CSU IF and the ISI IF, which indicates usage over the entire CSU community is inversely related to general citation impact.\\

The observed negative correlations between the CSU UIF and ISI IF run counter to previous findings. In fact, \citeA{earlie:brody2006} and \citeA{altern:bollen2005} report positive correlations between usage and citation rates. However, the services that recorded this usage, namely the UK arXiv mirror and the LANL Research Library systems, mostly accommodate a community of scholars in computer science and physics. The CSU community for wich usage was recorded is composed of a mix of students, faculty, staff and others, focused on a variety of science and social science domains. It can be speculated that both the nature of the CSU library collection as well as the CSU community that uses it jointly contributed to the negative correlations between CSU UIF and ISI IF values.\\

However, positive as well as negative CSU UIF vs. ISI IF correlations were observed for specific scholarly disciplines. In addition, a comparison of the relative sizes of the undergraduate and graduate communities at CSU to the correlations of CSU UIF vs. ISI IF values within specific disciplines, suggested that the size of the graduate community (students and faculty) relative to that of the undergraduate community within a discipline could be related to the magnitude of the observed CSU UIF vs. ISI IF correlations. The tentative linear relationship that was observed between the ratio of graduate to undergraduate enrollment and CSU UIF vs. ISI IF correlations raises the possibility that applications of usage data can take into account demographic data to extract different facets of impact. We must however caution that the latter observations are based on only those 4 disciplines for which significant or marginally significant CSU UIF vs. ISI IF correlations were observed. Future research could focus on validating these tentative results for a larger number of disciplines.\\

In Section \ref{sampleconsiderations} we distinguished two factors that shape metric-based assessments of scholarly impact, namely the formal definition of a metric  and the sample that it has been applied to.
Although the UIF has been defined to mimic the IF, the CSU UIF and ISI IF rankings in this manuscript have been generated for very different samples of the scholarly community. The ISI IF rests on citation data collected for a set of ISI-selected journals; its rankings therefore express the {\em global} community of all scholarly authors publishing in those journals. The CSU usage data on the other hand reflects the characteristics of the {\em local} CSU academic community that comprises a mix of students and faculty among others. It can therefore be considered at the same time more {\em diverse} than the ISI-defined sample in terms of its composition, yet more limited in terms of its {\em span} since it applies to CSU users only.\\

We envision three future paths along which usage-based metrics such as the UIF can be developed. These paths are not mutually exclusive and are related to the issues mentioned in the introduction.\\

The first path is one in which attempts are undertaken to mimic the properties of the ISI IF on the basis of usage data. This requires the aggregation of a meaningful, representative sample of the scholarly community, similar in span to the ISI IF sample, and efforts to compensate for the increased diversity of the usage data sample, e.g.~excluding all agents that are not scholarly authors and taking into account particular discipline-specific demographics and preferences. This article has provided an initial exploration of the second issue, whereas the architecture described in \cite{archit:bollen2006} may offer at least a technical solution to the first issue. Questions remain as to how one can create a truly representative usage sample of the global scholarly community.\\

The second path along which usage-based metrics of scholarly status can be developed is focused on leveraging the greater diversity (in terms of agents and community characteristics) that usage data generally engenders. This path may still require the aggregation of a meaningful, representative sample of the scholarly community, but its assessment of scholarly impact specifically leverages sample diversity to assess the many different facets of impact as they exist in the scholarly community.  Indeed, one could argue that an article that is often read by a majority of students, yet is seldom cited by scholars in this field, nevertheless has considerable scholarly impact. In fact, on the basis of sufficiently detailed usage data, impact could be separately assessed for any subset of the scholarly community including undergraduate and graduate students, research faculty, lecturers and the public at large.\\

Finally, where only local usage data is collected, there is still particular value in being able to determine local impact rankings which correspond to the preferences and characteristics of specific communities such as CSU. The CSU UIF generated in this article may not be globally applicable, but offers CSU administrators an interesting perspective on what is valued in their community.  Our analysis demonstrates that considerable, yet locally meaningful deviations can occur between impact as it is perceived by particular scholarly disciplines and the ISI IF. Such deviations are not problematic, but offer considerable possibilities to optimize local information services and adopt policies to accommodate the preferences of local communities.\\

Many issues remain to be addressed in future research on this topic. The Andrew W. Mellon foundation has awarded a grant to our team to investigate a range of issues related to the definition of usage-based metrics of scholarly impact. The funded project, named MESUR\footnote{\url{http://www.mesur.org/}}, aims to construct a large-scale model of the scholarly community which merges usage and bibliographic data to support the definition and validation of a range of usage-based metrics of scholarly status. This paper describes our first explorations in this research area.


\begin{thebibliography}{}

\bibitem[\protect\BCAY{{Analytic Studies Division}}{{Analytic Studies
  Division}}{2004}]{csu:stats2004}
{Analytic Studies Division}.
\newblock{}\BBOP{}2004\BBCP{}.
\newblock{}\Bem{Statistical abstract 2004-2005}\ (\BTR{}).
\newblock{}California State University.

\bibitem[\protect\BCAY{Bollen \BBA{} Luce}{Bollen \BBA{}
  Luce}{2002}]{evalua:bollen2002}
Bollen, J.\BCBT{} \BBA{} Luce, R.
\newblock{}\BBOP{}2002\BBCP{}.
\newblock{}\BBOQ{}Evaluation of digital library impact and user communities by
  analysis of usage patterns.\BBCQ{}
\newblock{}\Bem{D-Lib Magazine}, \Bem{8}(6).

\bibitem[\protect\BCAY{Bollen, Luce, Vemulapalli\BCBL{} \BBA{} Xu}{Bollen
  \BOthers{}}{2003}]{detect:bollen2003}
Bollen, J., Luce, R., Vemulapalli, S.\BCBL{} \BBA{} Xu, W.
\newblock{}\BBOP{}2003\BBCP{}.
\newblock{}\BBOQ{}Detecting research trends in digital library
  readership.\BBCQ{}
\newblock{}In \Bem{Proceedings of the seventh {E}uropean {C}onference on
  {D}igital {L}ibraries ({LNCS} 2769)}\ (\BPGS\ 24--28).
\newblock{}Trondheim, Norway: Springer-Verlag.

\bibitem[\protect\BCAY{Bollen \BBA{} {Van de Sompel}}{Bollen \BBA{} {Van de
  Sompel}}{2006\protect\BCnt{1}}]{mappin:bollen2006}
Bollen, J.\BCBT{} \BBA{} {Van de Sompel}, H.
\newblock{}\BBOP{}2006\protect\BCnt{1}\BBCP{}.
\newblock{}\BBOQ{}Mapping the structure of science through usage.\BBCQ{}
\newblock{}\Bem{Scientometrics}, \Bem{69}(2).

\bibitem[\protect\BCAY{Bollen \BBA{} {Van de Sompel}}{Bollen \BBA{} {Van de
  Sompel}}{2006\protect\BCnt{2}}]{archit:bollen2006}
Bollen, J.\BCBT{} \BBA{} {Van de Sompel}, H.
\newblock{}\BBOP{}2006\protect\BCnt{2}\BBCP{}.
\newblock{}\BBOQ{}An architecture for the aggregation and analysis of scholarly
  usage data.\BBCQ{}
\newblock{}In \Bem{{J}oint {C}onference on {D}igital {L}ibraries ({JCDL}2006)}\
  (\BPGS\ 298--307).
\newblock{}Chapel Hill, NC.

\bibitem[\protect\BCAY{Bollen, {Van de Sompel}, Smith\BCBL{} \BBA{}
  Luce}{Bollen \BOthers{}}{2005}]{altern:bollen2005}
Bollen, J., {Van de Sompel}, H., Smith, J.\BCBL{} \BBA{} Luce, R.
\newblock{}\BBOP{}2005\BBCP{}.
\newblock{}\BBOQ{}Toward alternative metrics of journal impact: a comparison of
  download and citation data.\BBCQ{}
\newblock{}\Bem{Information Processing and Management}, \Bem{41}(6),
  1419--1440.

\bibitem[\protect\BCAY{Brody, Harnad\BCBL{} \BBA{} Carr}{Brody
  \BOthers{}}{2006}]{earlie:brody2006}
Brody, T., Harnad, S.\BCBL{} \BBA{} Carr, L.
\newblock{}\BBOP{}2006\BBCP{}.
\newblock{}\BBOQ{}Earlier web usage statistics as predictors of later citation
  impact.\BBCQ{}
\newblock{}\Bem{Journal of the American Society for Information Science and
  Technology}, \Bem{57}(8), 1060 -- 1072.

\bibitem[\protect\BCAY{Darmoni, Roussel, Benichou, Thirion\BCBL{} \BBA{}
  Pinhas}{Darmoni \BOthers{}}{2002}]{readin:darmoni2002}
Darmoni, S.~J., Roussel, F., Benichou, J., Thirion, B.\BCBL{} \BBA{} Pinhas, N.
\newblock{}\BBOP{}2002\BBCP{}.
\newblock{}\BBOQ{}Reading factor: a new bibliometric criterion for managing
  digital libraries.\BBCQ{}
\newblock{}\Bem{Journal of the Medical Library Association}, \Bem{90}(3),
  323--327.

\bibitem[\protect\BCAY{Gallagher, Bauer\BCBL{} \BBA{} Dollar}{Gallagher
  \BOthers{}}{2005}]{eviden:gallagher2005}
Gallagher, J., Bauer, K.\BCBL{} \BBA{} Dollar, D.~M.
\newblock{}\BBOP{}2005\BBCP{}.
\newblock{}\BBOQ{}Evidence-based librarianship: Utilizing data from all
  available sources to make judicious print cancellation decisions.\BBCQ{}
\newblock{}\Bem{Library Collections, Acquisitions and Technical Services},
  \Bem{29}.

\bibitem[\protect\BCAY{Garfield}{Garfield}{1979}]{citati:garfield1979}
Garfield, E.
\newblock{}\BBOP{}1979\BBCP{}.
\newblock{}\Bem{Citation indexing: Its theory and application in science,
  technology, and humanities.}
\newblock{}New York: John Wiley and Sons.

\bibitem[\protect\BCAY{King, Tenopir\BCBL{} \BBA{} Clarke}{King
  \BOthers{}}{2006}]{measur:king2006}
King, D.~W., Tenopir, C.\BCBL{} \BBA{} Clarke, M.
\newblock{}\BBOP{}2006\BBCP{}.
\newblock{}\BBOQ{}Measuring total reading of journal articles.\BBCQ{}
\newblock{}\Bem{{D}-{L}ib {M}agazine}, \Bem{12}(10).

\bibitem[\protect\BCAY{Kurtz \BOthers{}}{Kurtz
  \BOthers{}}{2004\protect\BCnt{1}}]{bibliom:kurtz2004}
Kurtz, M.~J., Eichhorn, G., Accomazzi, A., Grant, C.~S., Demleitner, M.\BCBL{}
  \BBA{} Murray, S.~S.
\newblock{}\BBOP{}2004\protect\BCnt{1}\BBCP{}.
\newblock{}\BBOQ{}The bibliometric properties of article readership
  information.\BBCQ{}
\newblock{}\Bem{{JASIST}}, \Bem{56}(2), 111--128.

\bibitem[\protect\BCAY{Kurtz \BOthers{}}{Kurtz
  \BOthers{}}{2004\protect\BCnt{2}}]{worldw:kurtz2004}
Kurtz, M.~J., Eichhorn, G., Accomazzi, A., Grant, C.~S., Demleitner, M.\BCBL{}
  \BBA{} Murray, S.~S.
\newblock{}\BBOP{}2004\protect\BCnt{2}\BBCP{}.
\newblock{}\BBOQ{}Worldwide use and impact of the {NASA} {A}strophysics {D}ata
  {S}ystem digital library.\BBCQ{}
\newblock{}\Bem{{JASIST}}, \Bem{56}(1), 36--45.

\bibitem[\protect\BCAY{{MacRoberts} \BBA{} {MacRoberts}}{{MacRoberts} \BBA{}
  {MacRoberts}}{1989}]{proble:macroberts1989}
{MacRoberts}, M.~H.\BCBT{} \BBA{} {MacRoberts}, B.~R.
\newblock{}\BBOP{}1989\BBCP{}.
\newblock{}\BBOQ{}Problems of citation analysis: A critical review.\BBCQ{}
\newblock{}\Bem{Journal of the American Society for Information Science},
  \Bem{40}(5), 342--349.

\bibitem[\protect\BCAY{Mc{D}onald}{Mc{D}onald}{2006}]{unders:mcdonald2006}
Mc{D}onald, J.~D.
\newblock{}\BBOP{}2006\BBCP{}.
\newblock{}\BBOQ{}Understanding online journal usage: A statistical analysis of
  citation and use.\BBCQ{}
\newblock{}\Bem{Journal of the American Society for Information Science and
  Technology}, \Bem{57}(13).

\bibitem[\protect\BCAY{{Van de Sompel}}{{Van de
  Sompel}}{1999\protect\BCnt{1}}]{refere1:vandesompel1999}
{Van de Sompel}, H.
\newblock{}\BBOP{}1999\protect\BCnt{1}\BBCP{}.
\newblock{}\BBOQ{}Reference linking in a hybrid library environment (i).\BBCQ{}
\newblock{}\Bem{D-Lib Magazine}, \Bem{5}(4).

\bibitem[\protect\BCAY{{Van de Sompel}}{{Van de
  Sompel}}{1999\protect\BCnt{2}}]{refere2:vandesompel1999}
{Van de Sompel}, H.
\newblock{}\BBOP{}1999\protect\BCnt{2}\BBCP{}.
\newblock{}\BBOQ{}Reference linking in a hybrid library environment
  (ii).\BBCQ{}
\newblock{}\Bem{D-Lib Magazine}, \Bem{5}(4).

\bibitem[\protect\BCAY{{Van de Sompel} \BBA{} Beit-Arie}{{Van de Sompel} \BBA{}
  Beit-Arie}{2001}]{openli:vandesompel2001}
{Van de Sompel}, H.\BCBT{} \BBA{} Beit-Arie, O.
\newblock{}\BBOP{}2001\BBCP{}.
\newblock{}\BBOQ{}Open linking in the scholarly information environment using
  the {OpenURL} framework.\BBCQ{}
\newblock{}\Bem{D-Lib Magazine}, \Bem{7}(3).

\end{thebibliography}

\section*{Acknowledgements} We thank the Andrew W. Mellon Foundation for supporting this research. We also thank Marko A. Rodriguez for proofreading the earlier versions of this manuscript and Joan Smith at the Department of Computer Science at Old Dominion University for producing the raw citation data on which parts of this analysis are based. 

\onecolumn

\section*{Appendix}

\begin{table}[h!]
\begin{footnotesize}
\begin{center}
\begin{tabular}{||p{16cm}||}
\hline\hline
{\bf Agriculture and Natural Resources:}
AD (AGRICULTURE, DAIRY \& ANIMAL SCIENCE), AE (AGRICULTURAL ENGINEERING, AF (AGRICULTURAL ECONOMICS \& POLICY), AH (AGRICULTURE, MULTIDISCIPLINARY), XE (AGRICULTURE, SOIL SCIENCE) \\\hline\hline

{\bf Architecture and Environmental Design:}
IH  (ENGINEERING, ENVIRONMENTAL), JA (ENVIRONMENTAL SCIENCES, NE (PUBLIC, ENVIRONMENTAL \& OCCUPATIONAL HEALTH), JB (ENVIRONMENTAL STUDIES)\\\hline

{\bf Area Studies:}
BM (AREA STUDIES)\\\hline

{\bf Biological Sciences:}
CQ (BIOCHEMISTRY \& MOLECULAR BIOLOGY, CU (BIOLOGY), DB (BIOTECHNOLOGY \& APPLIED MICROBIOLOGY), DR (CELL BIOLOGY), HT (EVOLUTIONARY BIOLOGY), HY (DEVELOPMENTAL BIOLOGY), PI (MARINE \& FRESHWATER BIOLOGY), QU (MICROBIOLOGY), WF (REPRODUCTIVE BIOLOGY), BV (PSYCHOLOGY, BIOLOGICAL) \\\hline

{\bf Business and Management:}
DI (BUSINESS), DK (BUSINESS, FINANCE), PE (OPERATIONS RESEARCH \& MANAGEMENT SCIENCE), PC (MANAGEMENT) \\\hline\hline

{\bf Communications:}
YE (TELECOMMUNICATIONS, EU (COMMUNICATION) \\\hline

{\bf Computer and Information Sciences:}
EP (COMPUTER SCIENCE, ARTIFICIAL INTELLIGENCE), ER (COMPUTER SCIENCE, CYBERNETICS), ES (COMPUTER SCIENCE, HARDWARE \& ARCHITECTURE, ET (COMPUTER SCIENCE, INFORMATION SYSTEMS), EV (COMPUTER SCIENCE, INTERDISCIPLINARY APPLICATIONS), EW (COMPUTER SCIENCE, SOFTWARE ENGINEERING), EX (COMPUTER SCIENCE, THEORY \& METHODS), ET (COMPUTER SCIENCE, INFORMATION SYSTEMS), PT (MEDICAL INFORMATICS), NU (INFORMATION SCIENCE \& LIBRARY SCIENCE)\\\hline

{\bf Education:}
HB (EDUCATION, SCIENTIFIC DISCIPLINES), HA (EDUCATION \& EDUCATIONAL RESEARCH), HE (EDUCATION, SPECIAL), HI (PSYCHOLOGY, EDUCATIONAL)\\\hline

{\bf Engineering:}
AE (AGRICULTURAL ENGINEERING), AI (ENGINEERING, AEROSPACE), EW (COMPUTER SCIENCE, SOFTWARE ENGINEERING), IF (ENGINEERING, MULTIDISCIPLINARY), IG (ENGINEERING, BIOMEDICAL), IH (ENGINEERING, ENVIRONMENTAL), II (ENGINEERING, CHEMICAL), IJ (ENGINEERING, INDUSTRIAL), IK (ENGINEERING, MANUFACTURING, IL (ENGINEERING, MARINE), IM (ENGINEERING, CIVIL), IO (ENGINEERING, OCEAN), IP (ENGINEERING, PETROLEUM), IQ  (ENGINEERING, ELECTRICAL \& ELECTRONIC), IU (ENGINEERING, MECHANICAL), IX (ENGINEERING, GEOLOGICAL), PZ  (METALLURGY \& METALLURGICAL ENGINEERING)\\\hline

{\bf Fine and Applied Arts:}
No results \\\hline

{\bf Foreign Languages:}
No results\\\hline

{\bf Health Professions:}
HL (HEALTH CARE SCIENCES \& SERVICES), NE (PUBLIC, ENVIRONMENTAL \& OCCUPATIONAL HEALTH),  LQ (HEALTH POLICY AND SERVICES)\\\hline

{\bf Home Economics:}
No results \\\hline

{\bf Interdisciplinary Studies:}
EV (COMPUTER SCIENCE, INTERDISCIPLINARY APPLICATIONS, PO (MATHEMATICS, INTERDISCIPLINARY APPLICATIONS), WU (SOCIAL SCIENCES, INTERDISCIPLINARY)\\\hline

{\bf Letters:}
No results\\\hline

{\bf Library:}
NU (INFORMATION SCIENCE \& LIBRARY SCIENCE)\\\hline

{\bf Mathematics:}
PN (MATHEMATICS, APPLIED), PO (MATHEMATICS, INTERDISCIPLINARY APPLICATIONS), PQ (MATHEMATICS)\\\hline

{\bf Physical Sciences:}
UB (PHYSICS, APPLIED), UF (PHYSICS, FLUIDS \& PLASMAS), UH (PHYSICS, ATOMIC, MOLECULAR \& CHEMICAL), UI (PHYSICS, MULTIDISCIPLINARY), UK (PHYSICS, CONDENSED MATTER)\\\hline

{\bf Psychology:}
VI (PSYCHOLOGY), BV (PSYCHOLOGY, BIOLOGICAL), EQ (PSYCHOLOGY, CLINICAL), HI (PSYCHOLOGY, EDUCATIONAL), , MY (PSYCHOLOGY, DEVELOPMENTAL), NQ (PSYCHOLOGY, APPLIED), VJ (PSYCHOLOGY, MULTIDISCIPLINARY),  VP (PSYCHOLOGY, PSYCHOANALYSIS), VS (PSYCHOLOGY, MATHEMATICAL), VX (PSYCHOLOGY, EXPERIMENTAL), WQ (PSYCHOLOGY, SOCIAL)\\\hline

{\bf Public Affairs:}
NE (PUBLIC, ENVIRONMENTAL \& OCCUPATIONAL HEALTH), VM (PUBLIC ADMINISTRATION)|\\\hline

{\bf Social Sciences:}
PS (SOCIAL SCIENCES, MATHEMATICAL METHODS), WU (SOCIAL SCIENCES, INTERDISCIPLINARY), WV (SOCIAL SCIENCES, BIOMEDICAL)\\\hline\hline

\end{tabular}
\end{center}
\end{footnotesize}
\caption{\label{categories} ISI journal classification codes for CSU disciplines listed in Table \ref{enroll}.}
\end{table}

\end{document}